\documentclass[]{jfm}
\usepackage{}
\usepackage{graphicx}
\usepackage{newtxtext}
\usepackage{float}
\usepackage{newtxmath}
\usepackage{natbib}
\usepackage{hyperref}
\usepackage{verbatim} % Add this line

\usepackage[section]{placeins}
\hypersetup{
    colorlinks = true,
    urlcolor   = blue,
    citecolor  = blue,
}

\newcommand{\RomanNumeralCaps}[1]
\title{Role of particle volume fraction on particulate suspension droplet evolution, transition and Hysteresis}%Experimental studies on Transition Dynamics in Particle Laden Dripping-jetting  and Hysteresis effects}
\author{Kishorkumar Sarva\aff{1}
  \corresp{\email{kishorsarva@iisc.ac.in}}
 }

\affiliation{\aff{1} Interdisciplinary Center for Energy Research (ICER), IISc Bangalore, 560012, India 
 }

\begin{document}
\maketitle

\begin{abstract}

We study the transitional dynamics of the non-Brownian particulate Newtonian liquid jet for different particle volume fractions ($\phi$). We focus on the influence of particle volume fraction on the critical inflow velocity at which the dripping-jetting (i.e., dripping to jetting and jetting to dripping) transition occurs for the ratio of the nozzle diameter to the particle diameter ($D_n/D_p$=20). The experiments were conducted by increasing (forward sweep) and decreasing (reverse sweep) the flow rate. These experiments were repeated for different volume fractions. We observe, with an increase in particle volume fraction, the transition from the dripping to the jetting regime occurs through a chaotic dripping regime. With an increase in the particle volume fraction, the jetting regime has occurred at early flow rates during dripping to jetting transition (in forward sweep), and the jetting to dripping transition (reverse sweep) occurred at a lower flow rate than the forward sweep. The particle volume fraction impacts the hysteresis observed for the Newtonian fluid. Due to the changes in the critical flow rate where transition occur, the widening of the hysteresis loop of flow rate with the pinchoff length is observed. The transition from dripping to jetting is observed to have the recurrent escape of the pinchoff mechanism as the jet length changes, influencing the droplet size distribution. The frequency of droplet pinchoff and droplet size have decreased as the particle volume fraction has increased. As the particle volume fraction increases, the size distribution between the dripping and jetting regimes decreases.
   
\end{abstract}

%\begin{keywords}
%Authors should not enter keywords on the manuscript, as these must be chosen by the author during the online submission process and will then be added during the typesetting process (see \href{https://www.cambridge.org/core/journals/journal-of-fluid-mechanics/information/list-of-keywords}{Keyword PDF} for the full list).  Other classifications will be added at the same time.
%\end{keywords}

\section{Introduction}

%{\bf 
%\begin{enumerate}
%    \item Effect of particle number density for a given particle radius on a)Breakup length b) Hysteresis c)Droplet size  
%\end{enumerate}
%}

Particle-laden fluids are oft encountered in propellants, biological fluids, fluids in the food and pharmaceutical industries etc  (\cite{basaran2013nonstandard,zhang20213d},\cite{noh2022atomization, gvozdyakov2021improvement}, \cite{xu2014study}). Understanding the phenomenology of flow of such particle laden fluids is critical for appropriately designing processes in the industry. Traditionally, the regimes of flow of a Newtonian fluid form the backbone for these particle laden flows. The influence of particles is superposed on the response of Newtonian fluids. Especially important, are flow regimes through a nozzle or faucet.  
Frequently, in a typical Newtonian fluid,  the regimes of flow through a nozzle are divided into dripping , jetting  and atomization based on the flow rate. The physics and mechanics of these regimes, and the transitions between these are reasonably well established \cite{ kiyono1999dripping, clanet1999transition,ambravaneswaran2000theoretical,coullet2005hydrodynamical, lin1998drop}. These states of flow (henceforth called regimes) are observed by modulating the Weber number (ratio of inertia and surface tension). Details of the transient dynamics of flow from the dripping to jetting regimes and other critical phenomena such as hysterisis as we traverse through the flow regimes are all clearly elucidated elsewhere  \cite{sarva2024dynamics, clanet1999transition, umemura2014self}.\\ With addition of particles into the fluid, the influence on the dynamics of dripping, jetting and its transition change from the backbone Newtonian fluid are quite significant . The individual constituent particle morphology and the volume fraction of the particles both are known to influence the flow characteristics (\cite{furbank2007pendant,furbank2004experimental, thievenaz2022onset}). \\  
Further, flow of such suspensions is influenced by backbone fluid properties, and boundary conditions such as flow rate, and nozzle diameter(\cite{furbank2004experimental}). At low flow rates, the surface tension and the acceleration due to gravity govern the evolution of the suspension-pendent droplet (\cite{miskin2012droplet}). (\cite{bonnoit2012accelerated, bertrand2012dynamics, mathues2015capillary}) identified that particle rearrangements  within the droplet are a key mechanism for accelerated detachment when the neck diameter approaches the particle size (Bonnoit et al. 2012). The influence of particle volume fraction on drop formation and detachment is crucial, as higher volume fractions prevent long filament formation and influence drop shape evolution (Bertrand et al. 2012). Detailed quantitative analysis of capillary breakup dynamics reveals a novel deceleration stage before final breakup, correlating particle distribution with thinning dynamics (Mathues et al. 2015). The evolution of the neck with particles follows a power-law variation in time, specifically \( h \propto (t - t_0)^{2/3} \), due to particle-induced surface deformations (Miskin \& Jaeger, 2012). However, Zhao et al. (\cite{zhao2015inhomogeneity}) demonstrated that the neck evolution in suspension droplets consists of three different stages: initially behaving as a homogeneous fluid, then transitioning as particle concentration decreases in the pinch-off zone, and finally, the breakup is governed by the pure fluid background fluid (\cite{zhao2015inhomogeneity}. The initially thinning droplet behaves as a homogeneous fluid but accelerates as heterogeneity sets in, governed by local particle concentration fluctuations and defined coherence lengths dependent on particle size and volume fraction (\cite{thievenaz2022onset}). Due to the particle morphology and volume fraction, the suspension droplet shows different evolution stages compared to Newtonian fluid. A detailed treatise on the formation of droplets in a typical Newtonian fluid is presented elsewhere. 
 
The evolution of the droplet in a particle-laden flow begins with a homogeneous suspension in the syringe/tank, where particles are uniformly distributed. As the pendant drop emerges from the nozzle, initially, it retains uniform particle distribution \cite{miskin2012droplet}. This uniform distribution allows the suspension to be characterized by an effective viscosity, causing it to behave like a viscous Newtonian fluid (\cite{zhao2015inhomogeneity}. As the droplet diameter decreases, local particle arrangement becomes heterogeneous. As the droplet diameter approaches the particle diameter, leading to a reduction in the effective diameter \cite{bonnoit2012accelerated, bertrand2012dynamics}. In the final stage, the particle concentration decreases in the pinch-off zone, resulting in the suspension behaving similarly to the background fluid, with the breakup dynamics dictated by the properties of the fluid alone \cite{thievenaz2022onset}. This stratification of particles lowers the local viscosity in the pinch-off condition \cite{zhao2015inhomogeneity}. This phenomenon has been explained by studying the neck radius \cite{bonnoit2012accelerated, bertrand2012dynamics}. This process of droplet formation in a suspension is seen immaterial of the constituent particle volume fraction. Further, (\cite{lindner2015single}) have shown that a single particle in the filament will shift the point pinchoff and delay the time of pinchoff. However, in particle-laden droplets, the microstructure of the particle arrangement and their interaction with the free surface affect the evolution of the droplets (\cite{lindner2015single}) and droplet size \cite{furbank2004experimental}). However, at a high volume fraction, as the necking stage appears, the particle will interact with the interface and corrugate the interface. Recently, a new scaling was proposed at the transition stage (\cite{thievenaz2022onset}) and also estimated the effect of particle volume fraction and polydispersity on the scaling analysis. 

At higher flow rate, particle presence influences the fluid jet in two ways: the macroscopic properties of the suspension and the particle diameter scale interactions. \cite{furbank2004experimental} observed that increasing particle volume fraction leads to a reduction in jet length, an increase in the diameter of droplets, and disappearance of satellite droplets. They also observed that larger particles caused a more rapid transition from jetting to dripping due to disturbances created by the particles within the jet.  At higher particle volume fractions \(\phi= \)20\( \%\) \(\&\) 30\( \%\), the transition to jetting becomes smoother over a range of flow rates, eliminating the sudden transition seen in pure fluids. Further, smaller particles in lower concentrations  \(\phi= \) 5\( \%\) \(\&\) 10\( \%\) resulted in jet lengths nearly twice that of larger particles but only about half the length of the pure fluid jet, indicating that particle size significantly affects jet pinchoff length (\cite{furbank2004experimental, chateau2018pinch}).  The specific relationship for the breakup length in terms of effective viscosity ($\eta$), surface tension ($\sigma$), and extrusion velocity ($u_0$) is given by: \( L/L_0 = [1 - \tilde{d} (L/L_0)^{1/4}]^{4/3}\)
where \( L_0 = \left(\alpha \eta \sqrt{U u_0}/\sigma h_0 Z\right)^{4/3} Z\) and \( \tilde{d} = \left(2 U^2 L_0/u_0^2 Z\right)^{1/4} (nd/h_0)\). These equations show that particles shorten the jet length by introducing finite-size effects. Indicating, smaller particles exhibit behaviour closer to pure fluids, while larger particles cause more disturbances that influence jet breakup (\cite{chateau2018pinch}). \\ While \cite{furbank2004experimental} found the influence of volume fraction on dripping to jetting transition on breakup length and droplet size. The impact on the critical flow rate at which the dripping-jetting transition takes place, the droplet size distribution, and the influence of particle motion within the jet that modifies the micro-structural variations for particle stratification are not studied. 
 we investigate the role of particle volume fraction on the dripping-jetting transition dynamics, tip recoiling and their role in the droplet size distribution due to modulation of flow rate.  
 \section{Experimental Methodology}
\begin{figure} 
   \centering \captionsetup{width=\linewidth} 
  \captionsetup{justification=justified} 
  \includegraphics[scale=0.6]{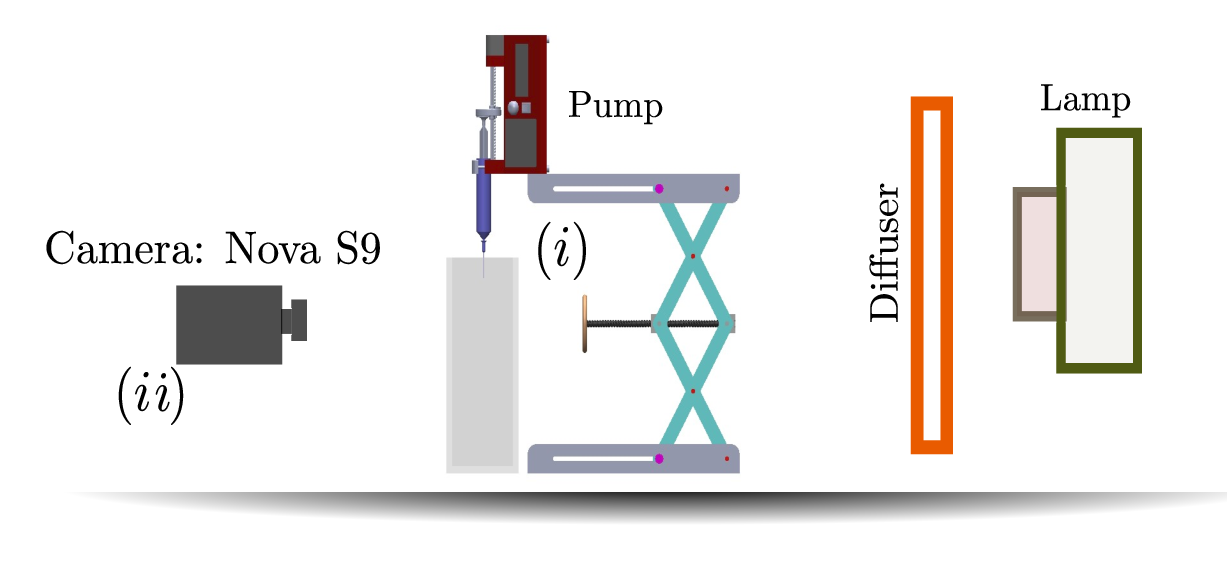}
  \captionof{figure}{\label{fig:testrig} The experimental setup for high-speed imaging is depicted. The various configuration modules are labelled in the diagram. $(i)$ indicated in the figure represents the droplet-generating module that includes an isolated chamber, a syringe pump, and needle assembly. $(ii)$ represents the shadowgraphy module with a high-speed camera, diffuser sheet, and lamp.} 
 \end{figure}

The schematic of the experimental setup is presented in figure \ref{fig:testrig}. The experiment was designed to delineate the influence of inflow conditions and particle volume fraction ($\phi$) on the transition dynamics between dripping and jetting of particle suspensions. The experimental setup includes a syringe pump (New Era Pump Systems, NE-1000) for precise control of flow rates attached to a syringe with a hypodermic needle, and a tank to collect the dispensed fluid and reduce external disturbances. The critical flow rate ($Q_c$) at which the transition to jetting occurs is a function of the nozzle diameter, fluid properties and particle characteristics \(Q_c( \rho, \sigma, \nu, g, \phi, D_p, D_n) \) (\cite{rubio2013thinnest, furbank2004experimental, sarva2022particle}). In order to study the influence of particle characteristics on the suspension dynamics, the experiments were conducted at the critical velocity for background fluid (22 $\%$ Glycerol solution) and the particles were chosen such that no jamming occurred through the experiment. (\cite{sarva2022particle}). \\
   \begin{table} 
  \begin{center}
\def~{\hphantom{0}}
  \begin{tabular}{lcccc}
    $Ka$ & $Oh$  & $U_m=\sqrt{We}$  &    $Bo$  &   $D_n/D_p$  \\ [3pt]
    0.0048 & 0.020  & 0.409-0.927 & 0.059  & 20    \\   
  \end{tabular}
  \caption{Experimental conditions of neutrally buoyant suspension prepared with aqueous glycerol solutions. At these system conditions, successive droplet evolution is observed at different volume fractions ($\phi$)  vary $0, 2 \%, 5 \%, 10 \%, 15 \%, 20 \%, 25 \%, 30 \%,  35 \%$ .}
  \label{tab:properties}
  \end{center}
\end{table}
To image the droplet evolution shadowgraphy was used. A high-speed camera (Photron Fastcam Nova SA9) with a field of view of 128 pixels x 1024 pixels was used to capture the droplet separation in the experiment. The field of view was illuminated by an LED source with a diffuser plate. The high-resolution images of the fluid jets at different particle volume fractions were processed to track the tip of the jet ($L$), pinchoff length ($L_b$), and droplet size ($D_d$). 
The pinchoff length of the jet ($L_b$) is determined by identifying the tip of the jet ($L$) in each frame using the Canny edge detection algorithm followed by dilatation. The edges of the jet and droplets were detected and tracked through the images. The minimum neck radius is calculated by finding the region of the drop along the horizontal in the dilated edge-detected image to find the thinnest part of the jet. The jet tip location is then marked on the frame to track the jet's evolution over time. The pinchoff location is detected by analyzing the jumps in the tip location. Contours of the droplet are outlined below the neck.  The diameter of each droplet ($D_d$) is calculated from the contour area ($A$) using \( D = \sqrt{4A/\pi} \). \\
 \begin{figure}
  \centering \captionsetup{width=\linewidth} 
  \captionsetup{justification=justified} 
  \includegraphics[scale=0.65]{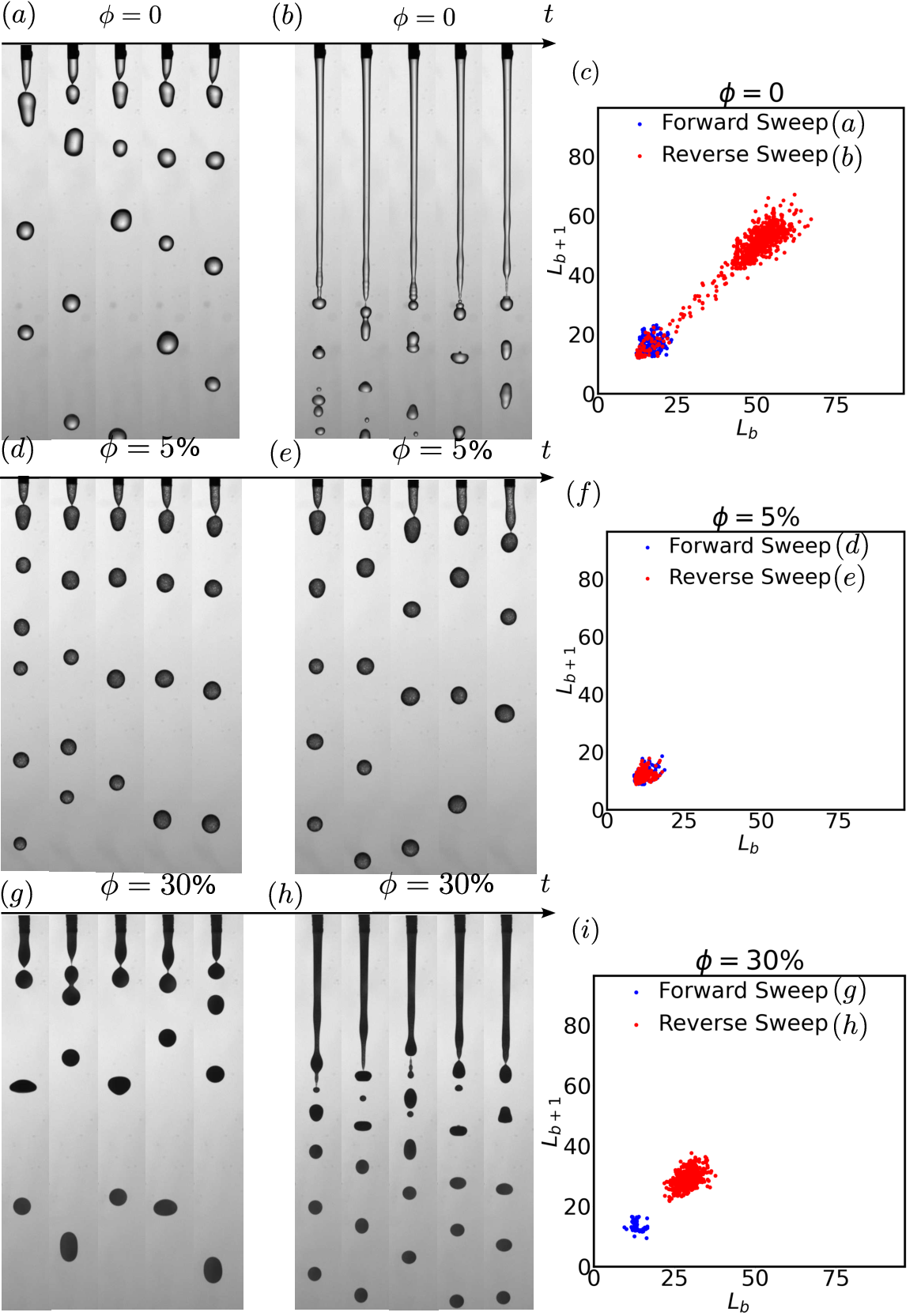}
  \captionof{figure}{Successive pinchoff images illustrate the impact of directional velocity variations, both ascending and descending, at $U_m= 0.889$ (Q=29 ml/m) for increasing volume fractions and their respective length return plots of pinchoff location. (a,b,c) Successive Newtonian droplet pinchoff in the dripping regime in ascending velocity and dripping-jetting transition in descending velocity. (d,e,f) Pinchoff profiles of dilute suspension $\phi=5 \%$, showing a periodic dripping regime. (g,h, i) Dripping regime in the ascending velocity and jetting regime in the descending velocity at $\phi=30 \%$.    \label{fig:flow_rate_direction} }  
% \caption{}  
 \end{figure}
The transition from dripping to jetting is governed by non-dimensional parameters used here i.e. Kaptiza Number ($Ka$), Weber number ($We$) and Bond Number ($Bo$). Wherein $Ka$ defines the influence of the fluid properties, $Ka = 3 \nu \left( \rho^3 g/\sigma^3 \right)^{1/4}$. It is a ratio of the viscous to capillary length scales $ 3(l_{\nu}/l_c)^{3/2}$, with $l_{\nu} = \nu^{2/3} g^{-1/2}$ representing the viscous length scale and  $l_c = \sqrt{\sigma/\rho g}$ is the capillary length scale. \\
Weber number or dimensionless inflow velocity ($U_m$), is the ratio of inertial to surface tension forces, defined as $U_m = U/U_c= \sqrt{We} = \sqrt{\rho U^2 R/\sigma_0}$. \\
Bond Number ($Bo$) is defined as $\rho g R_n^2/\sigma$, relating gravity to surface tension. \\
Lastly, the Ohnesorge Number ($Oh$) is the ratio between the viscous length scale ($l_{\mu} = \mu^2/\rho \sigma$) and the nozzle radius ($R_n$). The corresponding range of values is provided in table \ref{tab:properties}.\\
To study the influence of the direction of the inflow velocity  $U_m$ is varied every 5 seconds from 0.409 to 0.927. In the forward sweep, (i.e. increasing direction of flow rate) $U_m$ was increased from 0.409 to 0.927 and in the reverse sweep  (i.e. decreasing direction ) $U_m$ was reduced from 0.927 to 0.409. The influence of fluid properties of the particle-laden jet was studied for the diameter ratio $D_n/D_p= 20$ at different volume fractions of up to 35$\%$.
\section{Results and discussion}
\subsection{Dripping-jetting transition at fixed $D_n/D_p$ }
\begin{figure}
  \centering \captionsetup{width=\linewidth} 
  \captionsetup{justification=justified} 
  \includegraphics[scale=1]{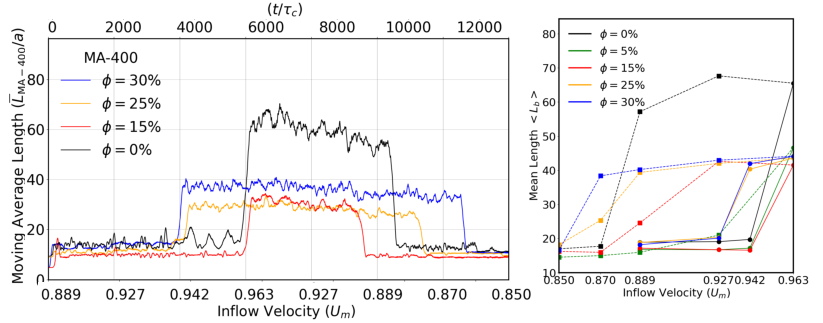}
  \captionof{figure}{(a) Depicts the Moving Average (MA-400) of the tip location of the fluid jet within the critical velocities over time at different volume fractions. The original real-time tip location is shown in Appendix figure  \ref{fig:tip_location} and tip tracking is shown in the movie-1. Each point on the moving curve represents the average of 400 consecutive data points, centered around each point to smooth the data. The figure shows the critical velocity at which the transition from dripping to jetting and jetting to dripping transition of the tip.  (b) Hysteresis curves for different volume fractions $\phi$ are shown between velocity and mean pinchoff length. The plot shows the mean pinch-off length as a function of velocity for both the increase and decrease of velocity. Each curve represents a different $\phi$, with different markers indicating the direction of the velocity: solid lines with filler markers for the increase and dashed lines with square markers decreasing the velocity.  The hysteresis loops widen with increasing $\phi$, suggesting particle loading intensifies the memory effect similar to the observations made to make use of hysterisis effect in the microfluid devices \cite{zhang20213d}. At lower velocities, mean lengths are almost stable. \label{fig:Benzier}   }  
 \end{figure}

\begin{figure} 
  \centering \captionsetup{width=\linewidth} 
  \captionsetup{justification=justified} 
  \includegraphics[scale=0.75]{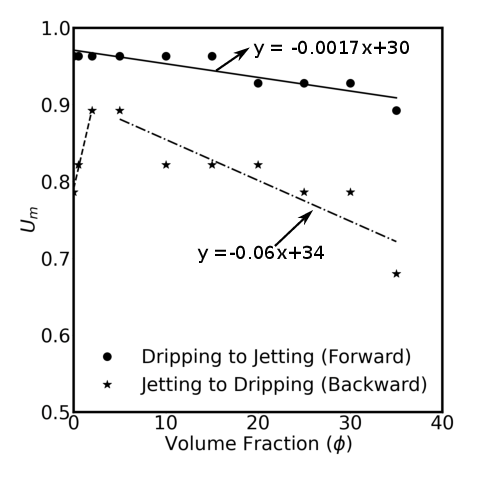}
  \captionof{figure}{The hysteresis observed in the critical velocity, where the transition from dripping to jetting and vice versa occurs as the particle volume fraction increases, is demonstrated. \label{fig:Bifurcation} }
 \end{figure}

\begin{figure} 
  \centering
  \captionsetup{width=\linewidth}
  \captionsetup{justification=justified} 
  \includegraphics[width=1\textwidth]{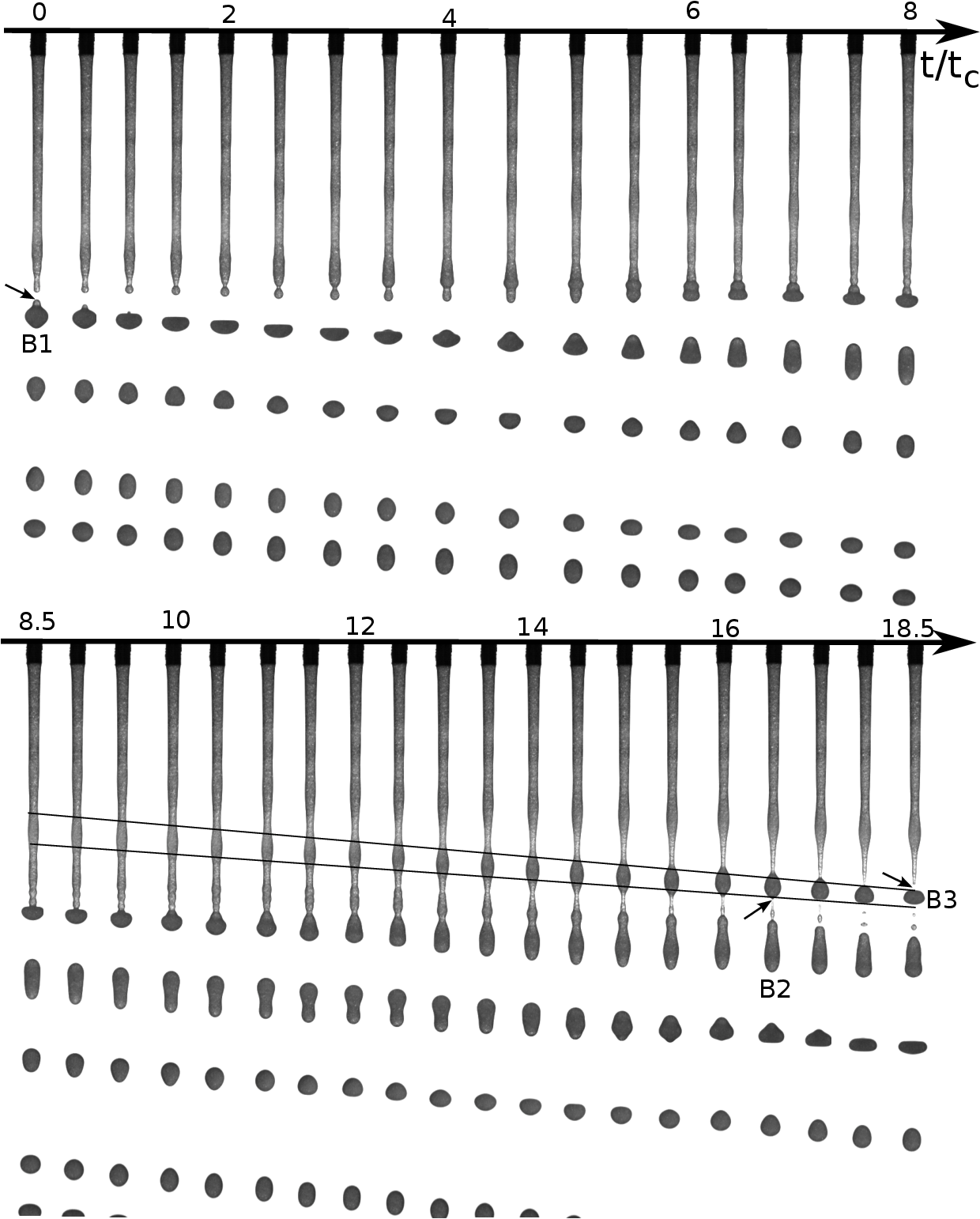}% Images in 100% size
  \caption{\label{fig:Breakup_modes} Tip evolution between successive pinchoff dynamics between B1 to B3 in $\phi=15 \%$ particulate suspension jet is shown in the figure at $U_m=0.927$. Following pinchoff B1, the tip will recoil, initiating capillary wave propagation upstream. The tip will bulge, oscillating as time progresses until 8 ms. During this period, capillary waves form crests and troughs.  However, Due to the escape of pinchoff, the tip will further elongate to form a prolate shape.  The particles will rearrange to accumulate more near the tip of the jet. Finally, the droplet will separate from the jet with a front pinchoff. The crest (Rayleigh instability) in the fluid jet shown from 8.5 ms is tracked between two lines to highlight the accumulation of particles compared to above, and below these lines leads to another pinchoff B3. Due to the particle size effect, the filament created between B2 and B3 leads to further breakup. }
  \end{figure}

First, we discuss the influence of the forward and reverse sweep of $U_m$ at different particle volume fractions for a fixed inflow velocity $U_m=0.889$. This discussion highlights the influence of direction from which $U_m$ is changed to arrive at $U_m=0.889$. Figure \ref{fig:flow_rate_direction} shows interface profiles and the corresponding Poincaré plots of pinchoff length ($L_b$) at $U_m=0.889$. While $U_m$ is constant, these interface profiles are shown For successive droplet pinchoff events at each volume fraction $\phi$. With change in the direction of $U_m$, the interface profiles show different droplet conditions. For example, figure \ref{fig:flow_rate_direction}a, shows the droplet pinch-offs of $\phi=0$ occur near the nozzle representing the dripping regime in the forward sweep and pinchoff of droplet far from the nozzle by forming a slender column representing the jetting regime in the reverse sweep of inflow velocity (see figure \ref{fig:flow_rate_direction}b) - which is referred to as hysteresis (\cite{umemura2014self}). The influence of particle volume fraction on these regime transitions are presented in the Poincaré plot of successive pinchoff lengths as shown in the figure \ref{fig:flow_rate_direction}c for $\phi=0$.\\
These Poincaré plot presents the nature of the successive pinchoff lengths $L_b$ allowing us to observe patterns that reflect the behaviour of successive pinchoff length $L_b$ (\cite{sartorelli1994crisis, kiyono1999dripping}). We can distinguish the patterns formed in the dripping and dripping-jetting transition in figure \ref{fig:flow_rate_direction}(c). This figure highlights the significant change in successive pinchoff length $L_b$, indicating the difference between the dripping regime and dripping-jetting transition. The cluster of points in blue colour in \ref{fig:flow_rate_direction}(c) in the dripping regime do not follow a clear trend, reflecting the irregular nature of pinchoff events. In the reverse sweep, a more scattered distribution with increasing length, shown in red, indicating a dripping to jetting transition. However, at a volume fraction of $\phi=5$, droplet interface profiles show dripping regime in both forward and reverse sweep in the figure \ref{fig:flow_rate_direction}d,e. Corresponding Poincaré plot \ref{fig:flow_rate_direction}f also show shorter pinchoff lengths ($L_b \sim 18$) indicating dripping. A similar cluster in the dripping regime is observed at $\phi=5 \%$ under both increasing and decreasing flow rates (see plot \ref{fig:flow_rate_direction}(f)). \\
At $\phi=30$, regime transition is observed as the direction of the $U_m$ sweep is changed in figure \ref{fig:flow_rate_direction}. We note in the reverse sweep of velocity (from $U_m=0.927$ to $U_m=0.409$), at $\phi=30\%$ jetting is observed in figure \ref{fig:flow_rate_direction}(h) at the same velocity ($U_m=0.889$). In the Poincaré plot in figure \ref{fig:flow_rate_direction}i, chaotic structure (u-shaped pattern) is observed in figure \ref{fig:flow_rate_direction}(i), akin to the structure observed of the dripping faucet of a Newtonian fluid in a backward period-doubling route to chaos by \cite{kiyono1999dripping}.\\
Further, the interface pinchoff profiles in both $\phi=0$ and $\phi=5$ exhibit axial asymmetry at the point of pinchoff, which reduces as the volume fraction increases, as shown for $\phi=30 \%$ in the figure \ref{fig:flow_rate_direction}(g,h). \\
However, From the Poincaré plots at different volume fractions, we observe that with an increase in the volume fraction, pinchoff length $L_b$ decreases, similar to the observations made by \cite{furbank2004experimental}. It is interesting to note that the tip length ($L$) and mean pinchoff length $<L_b>$ of the fluid jet are a function of the direction of the flow rate change (i.e., whether the flow rate is increasing or decreasing). This is presented as a function of $U_m$ in the figure \ref{fig:Benzier}.\\ 
 Plot \ref{fig:Benzier}a shows the moving average (MA-400 represents the average of 400 consecutive data points, centered around each point to smooth the data) of the tip length. This smoothing allowed  removing the noise and highlight trends of the tip over time, making it easier to identify critical velocities at which the transition from dripping to jetting and jetting to dripping occurs) and mean pinch-off length as a function of velocity for both the forward and reverse sweep of $U_m$. This is obtained by first performing experiments with the background fluid ($\phi=0 \%$) to see the effect of the velocity on the dripping-jetting (i.e. dripping to jetting and jetting to dripping) transition of the fluid jet by increasing and decreasing the velocity after every 5-second interval. 
 Later, the experiments were conducted at increasing volume fractions under similar conditions. The corresponding moving average of the tip location within the critical velocities over different volume fractions is shown in figure \ref{fig:Benzier}(a). In figure \ref{fig:Benzier}a, background fluid ($\phi=0$) shows chaotic length (no specific pattern) in the dripping regime below $U_m=0.963$, similar to the observations of \cite{clanet1999transition} for low viscosity fluids. As the velocity is decreased to 0.927 during the reverse sweep cycle, the moving average length of the jet decreases.\\
The transition from jetting to dripping is observed at the critical velocity $U_m= 0.889$. At 0.870, a stable moving average length (dripping regime) is observed during the reverse sweep cycle.\\ 
The influence of particle volume fraction on the critical velocity transitions are observed in both directions is shown for $\phi ($\%$) = 15, 25, 30$  in the plot \ref{fig:Benzier}(a). Figure shows the dripping-to-jetting transition as $\phi$ increases at lower velocities 0.942 for $\phi > 15 \%$. However, for dilute particle volume fraction at $\phi=15 \%$ (shown in the figure \ref{fig:Benzier}(a)), dripping-to-jetting transition at the same velocity as the $\phi=0 \%$. The transition occurrs with a significant reduction in tip length and a reduced number of droplet separations in the jetting-to-dripping transition seen at 0.889. The transition from jetting to dripping is observed at 0.870 for $\phi=25\%, 30 \%$, with the delayed transition for higher particle volume fractions. This altered appearance of the dripping-jetting transition is also reflected through their hysteresis during cycling of the flow rates. Figure \ref{fig:Benzier}(b) shows hysteresis curves for different volume fractions $\phi$ between velocity and mean pinchoff length. Each curve represents a different $\phi$, with different markers indicating the direction of the velocity: solid lines with filled markers for the forward sweep and dashed lines with square markers reverse sweep of the velocity (x-axis). The mean pinchoff length $<L_b>$ at different volume fractions shown in figure \ref{fig:Benzier}(b) indicates an increase in the $<L_b>$ in both forward and reverse directions indicating hysteresis. \\ 
  \begin{figure} 
  \centering \captionsetup{width=\linewidth} 
  \captionsetup{justification=justified} 
  \includegraphics[scale=0.65]{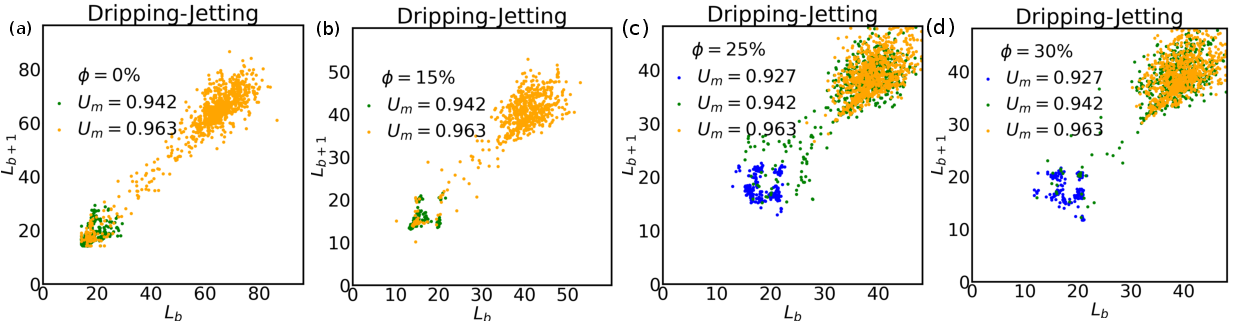}
  \captionof{figure}{ \label{fig:Poincaré_length} Within the critical velocity conditions, this figure displays Poincaré plots illustrating the breakup locations of droplets as the particle volume fraction increases. Each colour represents a different velocity, highlighting the transition from dripping to jetting as the velocity increases.}  
 \end{figure}
The variations in jet behavior are influenced by the $D_n/D_p$ ratio and particle volume fraction $\phi$ \cite{furbank2004experimental}. Suspension jets experience resistance to column thinning and particle interaction with the free interface \cite{furbank2004experimental}. In the experiments conducted here, we used $D_n/D_p = 20$ (small particle diameter), allowing us to perform experiments with up to $35\%$ volume fraction without clogging the nozzle. In the present experiments, we observe an increase in particle volume fraction stabilizes the fluid column for $D_n/D_p = 20$. The jet length reduces during the dripping to jetting transition with increased particle volume fraction, occurring at a lower velocity compared to $\phi=0$ due to increased effective viscosity \cite{furbank2004experimental, chateau2018pinch}. Furbank and Morris \cite{furbank2004experimental} attributed this reduced length to the increased resistance from the large particle volume fraction. 
\cite{eigenbrod2020effective} observed that, particle interaction with the interface will delay the decay of fluid slender jet.\\
In the dilute regime, particles move freely along the jet. Due to the moderate viscosity of the background fluids, capillary waves propagate long distances \cite{umemura2014self}. The pressure difference from capillary instability affects particle motion and causes partcle stratification along the jet, as observed from the jet shadowgraphy images (Figure \ref{fig:Breakup_modes}). The capillary instability creates a relative velocity between particles and fluid. This change in the local volume fraction, effects the the local viscosity, as seen in the capillary bridge experiment by \cite{mathues2015capillary}. In the dilute limit, isolated particles may interact with the interface. As the volume fraction increases, adjacent particles increasingly influence free particle motion, and particle diameter begins to affect jet dynamics, primarily due to clogging during jet pinchoff, where the jet interface will corrugate \cite{chateau2018pinch}. Such wavy interfaces were not observed under present experimental conditions for dilute regimes.\\
At the point of pinchoff, particles were observed to clog the neck, resulting in early breakup at higher $30\%$ volume fractions. Higher particle volume fractions alter the morphology of interface, where particle diameter also influences the evolution process. These observations suggest that increased effective viscosity due to higher volume fraction requires a greater velocity reduction for the jetting-to-dripping transition, as shown in Figure \ref{fig:Benzier} (a, b). For a dilute suspension at $5\%$, Figure \ref{fig:Benzier} shows a smaller mean length ($<L(\phi=5\%)> = 44$ mm) at every velocity compared to the fluid ($<L(\phi=0\%)> = 60$ mm) for $U_m = 0.963$. The hysteresis loops widen with increasing $\phi$.\\
Figure \ref{fig:Poincaré_length} presents the Poincaré plots for increasing volume fractions from $0\%$ to $30\%$. These plots illustrate the influence of velocity at each volume fraction on the nature of the pinchoff length $L_b$. The pinchoff lengths are extracted from each inflow velocity as the tip of the jet evolves with increasing velocity, as indicated in Figure \ref{fig:Benzier}(a). For $\phi=0$ shows dripping at $U_m=0.942$, dripping to jetting transition at $U_m=0.963$ in Figure \ref{fig:Poincaré_length}(a). The green point (representing $U_m=0.942$) form a relatively tight cluster, indicating a more stable and predictable regime. The orange points (representing $U_m=0.963$) are more dispersed, showing increased variability and indicating a transition to jetting. 
This transition includes 50 pinchoff events in the dripping regime, 30 droplet pinchoff events in the dripping to jetting transition, and, finally, 300 droplet pinchoff from the liquid jet in the jetting regime. For $\phi=15 \%$ (see  in figure \ref{fig:Poincaré_length}(b)) the green points ($U_m=0.942$) display a period-4 regime with intermittent behavior, characterized by clusters that repeat every four cycles but are irregular intervals. This intermittent periodic behavior was observed in the Newtonian dripping faucet experiments by \cite{sartorelli1994crisis}. The orange points ($U_m=0.963$) show a more dispersed pattern, suggesting a transition to jetting regime. During the transition, number of pinchoff events in the dripping regime (shorter length pinchoff) have decreased compared to $\phi=0$.\\
However, further increasing the volume fraction to $\phi=25\%$ and $\phi=30\%$, causes early appearance of dripping to jetting transition at 0.942 compared to lower volume fractions in the figures \ref{fig:Poincaré_length} c,d. The blue points at $u_m=0.927$, indicating a transition to a more chaotic dripping regime. At $U_m=0.942$ pinchoff length transition to jetting, and the number of pinchoff events in the dripping regime decreases as $\phi$ increases from 25 $\%$ to 30 $\%$. The orange points at $U_m=0.963$ exhibit jetting regime. Figure \ref{fig:Bifurcation} illustrates the hysteresis observed in the dripping-jetting transition as a function of velocity at different volume fractions represented in a $\phi$ - $U_m$ phase plot at a constant $Bo=0.102$. In the forward sweep (dripping to jetting transition), lower velocities are required for the transition at higher volume fractions, indicating stabilizing effect of volume fraction on the jet. In the backward sweep (jetting to dripping transition), the transition occurs at smaller velocities than the forward sweep as volume fraction increases, showing that higher volume fractions stabilizing effect, causing the transition to be delayed. However, at dilute limit, the jetting to dripping regime initially occured at higher velocities  upto $\phi=5 \%$ after which increasing volume fraction leads to delayed transition from dripping to jetting. This behavior highlights the influence of particle volume fraction and particle morphology effects on the transition dynamics, differing from the typical behavior observed in Newtonian fluid (\cite{Ambravaneswaran2004dripping, Subramani2006simplicity, rubio2018dripping}).  
\subsection{Droplet size distribution and driving mechanism}
To understand the influence of jet evolution on droplet size distribution, we examine the Poincaré plots at different volume fractions. However, during this evolution process, the mechanisms that govern the increase in tip length ($L$) and pinchoff length ($L_b$) and the droplet size distribution ($D_p$) were not discussed. Figure \ref{fig:Breakup_modes} highlights these mechanisms, such as the growing Rayleigh instability and capillary wave propagation due to tip recoiling and bulge formation in a stable fluid jet of $15 \%$ volume fraction. Figure \ref{fig:Breakup_modes} shows the time evolution of the fluid jet between three successive pinchoff conditions, namely B1, B2, and B3. Once the B1 droplet separates from the jet, two different mechanisms may occur: the tip retraction mechanism and the Rayleigh instability in the fluid jet. As droplet B1 separates from the fluid jet at $t/t_c=0$, the fluid jet retracts for a satellite droplet separation at $t/t_c=2$ that connects to the fluid jet with a particle-free neck. This particle-free neck stage was observed in the last stages of suspension droplet pinchoff in non-inertial droplet pinchoff conditions (\cite{miskin2012droplet, mathues2015capillary, lindner2015single, thievenaz2022onset}). However, before the droplet pinchoff, the neck enlarges to create a bulge at the tip. As time progresses, the bulge grows at the tip. As the time progress, the bulge shape oscillate while creating capillary wave propagation in the upstream direction. Shape oscillations of the tip lead to another escape of pinchoff at $t/t_c=12$. This escape of the pinchoff mechanism was observed during filament recoiling and dripping to jetting transition for Newtonian fluids (\cite{hoepffner2013recoil, sarva2024dynamics}). In the final stage of evolution, the oscillating tip (droplet) is connected to the jet with a particle-laden filament. Finally, Droplet B2 separates with a front pinchoff of the fluid jet along with a long particulate filament. This B2 droplet separation is due to the tip retraction mechanism. As the droplet separates from the liquid jet with a long filament, whose aspect ratio ($L/R$) influences satellite droplet formation \cite{notz2004dynamics}. Here, the filament's width is proportional to the particle's diameter; resulting in strong interaction between particle with the interface. Satellite droplets further break up into sub-satellite droplets after B2 separates from the jet in figure \ref{fig:Breakup_modes}. Simultaneously, the instability in slender jet shows an unstable wave as indicated with two lines in the figure \ref{fig:Breakup_modes} from $t/t_c=8.5$ due to the Rayleigh instability similar to the inviscid fluid jet during the jetting to dripping transition \cite{umemura2014self}. This wavelength grows to back-pinchoff (B3) from the jet at $t/t_c=18.3$.  \\  
\begin{figure} 
  \centering \captionsetup{width=1\linewidth} 
  \captionsetup{justification=justified} 
  \includegraphics[scale=0.6]{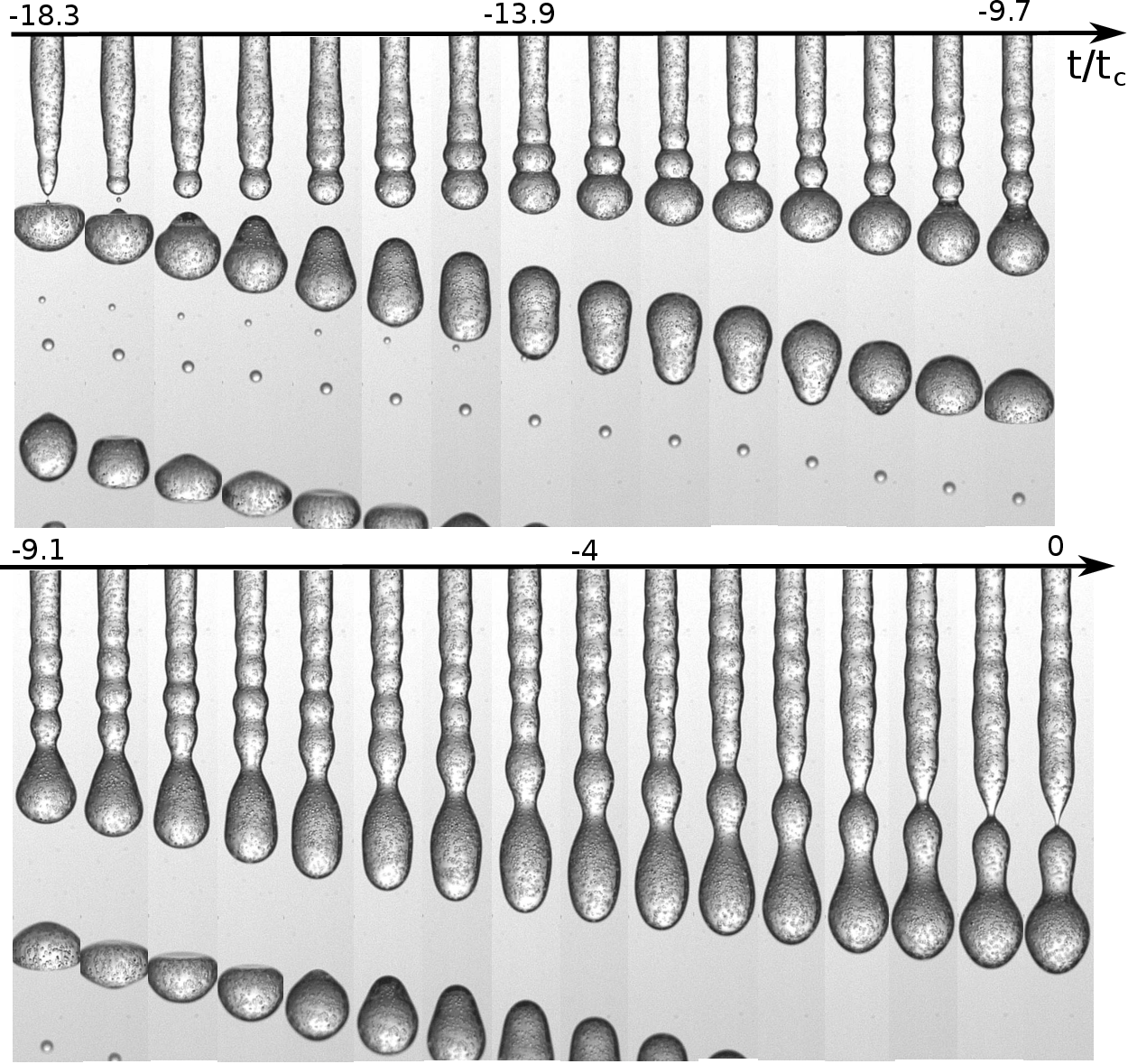}
  \captionof{figure}{\label{fig:Escape_of_pinchoff} Figure shows events at the tip of suspension jet. Such as tip recoil, capillary wave propagation, particle movement towards the tip bulge, and escape of pinchoff and pinchoff. The particle volume fraction in the droplet will increase during these events. The movie clip (Movie-3) illustrates the particle movement into the tip bulge from the escape of pinchoff, illustrating the relative velocity of particles from the tip movement. Here $\phi=5 \%$, $U_m= 0.886$.
  }  
 \end{figure} 
To further highlight the influence of the retraction mechanism on the droplet size distribution, figure \ref{fig:Escape_of_pinchoff} is shown at $\phi=5\%$. This figure highlights the tip retraction mechanism of droplet pinchoff at $t/t_c=0$. As the droplet separates at $t/t_c=-18.3$, the fluid jet tip retracts, forming a bulge at the tip and the appearance of the capillary wave travelling in an upward direction \cite{umemura2014self}. Due to these capillary waves, local velocity and pressure variations create a relative velocity between fluid and particles. The capillary waves interact \cite{safronov2021fast}, causing particle distribution variation and stratification within the fluid jet. During the evolution, the jet also shows the escape of the pinchoff mechanism at $t/t_c=-9.1$. This escape of the pinchoff event creates a vortex ring \cite{hoepffner2013recoil} and increases the local volume fraction within the droplet. The stratification of particle volume fraction and droplet influx into the droplet is shown in the supplementary video (Movie-3) in figure \ref{fig:Escape_of_pinchoff}.\\ \cite{safronov2021fast} highlights the  dampening effect of an increase in viscosity $\mu$ on capillary wave propagation in Newtonian liquid. The observed tip oscillations of the jet generate a propagating fast wave. Due to viscous dissipation, wave crest expands along with a decrease in the perturbation velocity. These results indicate that the interaction of waves generates a large wavelengths (k<1). \cite{safronov2021fast} models the propagation of axisymmetric perturbations on jet, using slender jet equation \cite{eggers2008physics}. The local perturbations for viscous Newtonian liquid jet was modeled by setting the tip (z=0) into motion with \(h(z=0,t)=sin( \omega t) \) condition to arrive at equation 3.1 (see details \cite{safronov2021fast}). This equation indicates the edge of the stationary jet periodically "swings", short waves with a spatial frequency  $k > 1$ are formed on the jet interface \cite{safronov2021fast}. The relationship between wave number ($k$), frequency ($\omega$), and the Ohnesorge number is  (\(Oh= \mu/\sqrt{\rho R_n \sigma}\)) discussed in their work. The equation is given by:\\
\begin{equation}
k = \pm \sqrt{\frac{1 - 6 \omega \text{Oh}}{2}} \left( 1 + \sqrt{1 + \frac{8 \omega^2}{(1 - 6 \omega \text{Oh})^2}} \right)
\end{equation}
\begin{equation}
k = \pm \sqrt{\frac{1 - 6 \omega \text{Oh} \left(1 - \frac{\phi}{\phi_c}\right)^{-2}}{2}} \left( 1 + \sqrt{1 + \frac{8 \omega^2}{\left(1 - 6 \omega \text{Oh} \left(1 - \frac{\phi}{\phi_c}\right)^{-2}\right)^2}} \right)
\end{equation}
  \begin{figure} 
  \centering \captionsetup{width=1\linewidth} 
  \captionsetup{justification=justified} 
  \includegraphics[scale=0.33]{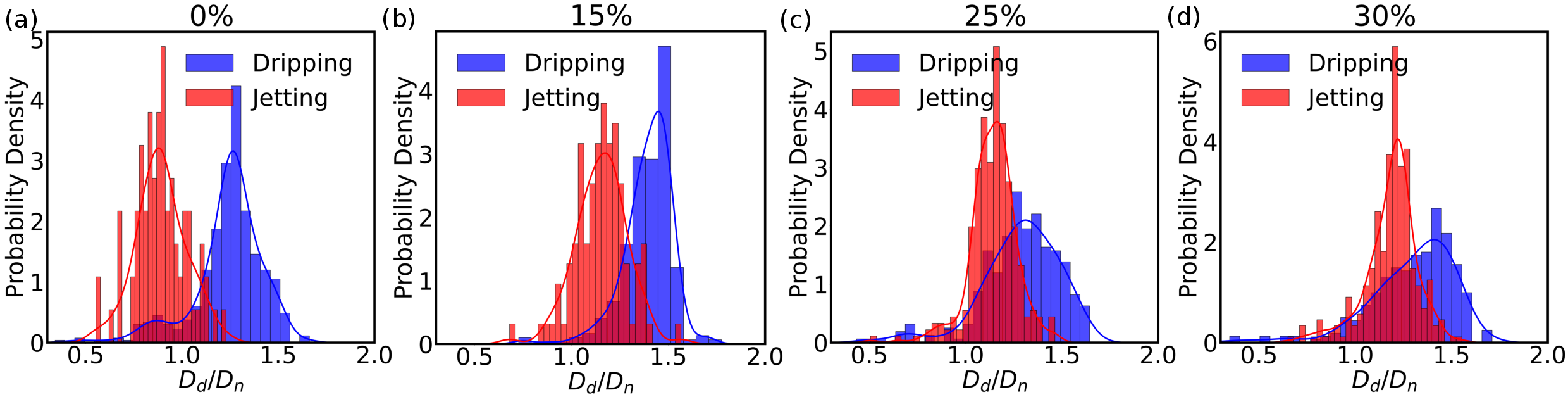}
  \captionof{figure}{\label{fig:hystorgram_dropsize} The histograms of droplet diameter distribution during dripping to jetting regime as flow rate and volume fraction change. The Kernal Distribution function shows the modes in droplet size distribution. The droplet size distribution is plotted from $U_m= 0.709$ (26 ml/m) to $U_m= 0.927$ (34 ml/m) flow rate in the dripping to jetting transition. The corresponding video is provided in the supplementary video movie-2.
  }  
  \end{figure}
The above equation 3.1 can be written in terms of $\phi$ by substituting Maron and Pierce relation, $\mu_{eff}= \mu (1- \phi/\phi_c)^{-2}$  \cite{maron1956application} into the suspension Ohneroge number defined as \(Oh_s= \mu_{eff}/\sqrt{\rho R_n \sigma}\) into equation 3.1 resulting in equation 3.2. From experimental observations, the volume fraction of the liquid jet is a function of axial location ($z$). This observation justifies to define Oh based on the local volume fraction similar to the equation derived by \cite{mathues2015capillary}. As the local volume fraction of the suspension increases, the lcaol effective viscosity increases, leading to a higher Oh. Consequently, a higher Oh results in greater damping of capillary waves, reducing $k$ and $\omega$, stabilising the jet tip oscillations. The damping effect due to higher viscosity may cause fewer droplet separation events, increasing the droplet size in the jetting regime. 
Figure \ref{fig:hystorgram_dropsize} shows histograms with Kernel Distribution Function (a kernel distribution function (KDF) is a non-parametric method to estimate the probability density function of a random variable) to highlight the dominant probability density of droplet size distribution for dripping (blue), dripping to jetting transition and jetting regime (red). Figure \ref{fig:hystorgram_dropsize}a shows droplet size distribution for Newtonian fluid ($\phi=0$). The droplet size distribution in dripping is clearly observed with dominating distribution $D_p/D_n \sim 0.8$ and $D_p/D_n \sim 1.2$ in the jetting regime. A comparable trend is observed upto $\phi=15 \%$ in figure \ref{fig:hystorgram_dropsize}b. As volume fraction further increases, the difference in droplet size distribution in both dripping and jetting regimes becomes narrower and more uniform at higher $\phi=25 \%$ $\&$ $30 \%$ in figure \ref{fig:hystorgram_dropsize}c,d. The jetting regime consistently shows a more concentrated distribution of droplet sizes compared to the dripping regime. The convergence of droplet size distributions in the dripping and jetting regimes with increasing volume fraction suggests a strong influence of volume fraction on droplet formation dynamics. 
\section{Conclusion}
We have studied the influence of particle volume fraction on the dripping-jetting transition regime and on hysteresis during the transition. Based on the pinchoff length and droplet volume, the regimes are identified as the dripping regime and the jetting regime. At low flow rates, droplet pinchoff occurs near the nozzle, and it is defined as a dripping regime. With an increase in flow rate, the pinchoff locations move far from the nozzle, drastically increasing the pinchoff length and reducing droplet size in the jetting regime. The transition is also path-dependent, due to which hysteresis is observed. Here, hysteresis is a function of flow rate and fluid properties. The particle volume fraction has an impact on this hysteresis. With an increase in the particle volume fraction, we have observed that the jetting regime has occurred at early flow rates, and the jetting to dripping transition has occurred at lower flow rates. This resulted in the widening of the hysteresis loop of flow rate with pinchoff length. It is also observed that with increasing flow rates, the dripping-to-jetting transition has occurred through a simple dripping-jetting transition. Here, we have observed that droplet size is influenced by the particle volume fraction. The distinct variation in droplet size between dripping and jetting is reduced with an increase in volume fraction. We observed this size distribution due to the escape of the pinchoff mechanism. It is observed that, with an increase in $\phi$, the influence of this mechanism is strongly observed, which will increase the droplet volume. It was also observed that, at the escape of pinchoff, particles will rush into the bulge formed at the tip, which may result in an increase in the particle volume fraction in the droplet. 
 %\begin{itemize}
%\item Experimental observations on dripping dynamics of Newtonian liquids and suspension are presentated.
%\item Increase in the flow rate, particle volume fraction results in change in dripping dynamics. 
%\item At large volume fraction and at small flow rate (1 ml/m), the breakup structure is modified from an end-pinchoff to mid pinchoff results in disappearance of satellite drop.  
%\item For moderate viscous liquid, with increase in particle volume fraction ($\phi$), the breakup structure modified from a $chaotic$ $dripping$ $->$ $periodic$ $->$ $Jetting$.  
%\item The breakup structure of viscous fluid quite different from the moderately viscous fluid in terms of filamnet length, jet length.
%\item With increase in the volume fraction in the jetting regime (29 ml/m), Breakup length is seen decreasing. 

% \end{itemize}      
 \appendix
 %  \begin{figure} 
%   \centering \captionsetup{width=\linewidth} 
%  \captionsetup{justification=justified} 
%  \includegraphics[scale=0.65]{suspension/viscosity.eps}
%  \captionof{figure}{ (a) viscosity of the suspensions with different particle volume fraction ($\phi$). (b) Relative effective viscosity of the suspension for different volume fractions.  } \label{fig:SHear_rheology} 
% \end{figure} 

 \begin{figure} 
   \centering \captionsetup{width=\linewidth} 
  \captionsetup{justification=justified} 
  \includegraphics[scale=1.4]{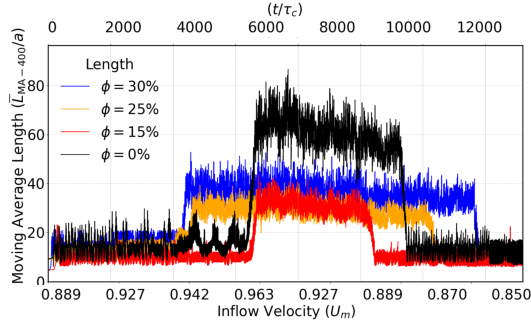}
  \captionof{figure}{ \label{fig:tip_location} Tracking the tip location with increase and decrease in the flow rate at each volume fraction.  } 
 
 \end{figure} 
 \bibliographystyle{jfm}
\bibliography{jfm}

@article{zhang20213d,
  title={3D printing with particles as feedstock materials},
  author={Zhang, Jun and Amini, Negin and Morton, David AV and Hapgood, Karen P},
  journal={Advanced Powder Technology},
  volume={32},
  number={9},
  pages={3324--3345},
  year={2021},
  publisher={Elsevier}
}

@article{notz2004dynamics,
  title={Dynamics and breakup of a contracting liquid filament},
  author={Notz, Patrick K and Basaran, Osman A},
  journal={Journal of Fluid Mechanics},
  volume={512},
  pages={223--256},
  year={2004},
  publisher={Cambridge University Press}
}

@article{rubio2018dripping,
  title={Dripping dynamics and transitions at high Bond numbers},
  author={Rubio-Rubio, Mariano and Taconet, Paloma and Sevilla, Alejandro},
  journal={International Journal of Multiphase Flow},
  volume={104},
  pages={206--213},
  year={2018},
  publisher={Elsevier}
}

@Article{rubio2013thinnest,
  title={On the thinnest steady threads obtained by gravitational stretching of capillary jets},
  author={Rubio-Rubio, Mariano and Sevilla, Alejandro and Gordillo, Jos{\'e} Manuel},
  journal={Journal of Fluid Mechanics},
  volume={729},
  pages={471--483},
  year={2013},
  publisher={Cambridge University Press}
}

@Article{clanet1999transition,
  title={Transition from dripping to jetting},
  author={Clanet, Christophe and Lasheras, Juan C},
  journal={Journal of fluid mechanics},
  volume={383},
  pages={307--326},
  year={1999},
  publisher={Cambridge University Press}
}

@inproceedings{sarva2022particle,
  title={Particle Filtration in Suspension Droplet Breakup},
  author={Sarva, Kishorkumar and murthy, Tejas G and Tomar, Gaurav},
  booktitle={Conference on Fluid Mechanics and Fluid Power},
  pages={387--394},
  year={2022},
  organization={Springer}
}

@article{sarva2024dynamics,
  title={Role of Dynamics in Dripping-Jetting Transition in Newtonian Fluids},
  author={Sarva, Kishorkumar and murthy, Tejas G and Tomar, Gaurav},
  journal={arXiv preprint arXiv: },
  year={2024}
}

@article{lin1998drop,
  title={Drop and spray formation from a liquid jet},
  author={Lin, Sung P and Reitz, Rolf D},
  journal={Annual review of fluid mechanics},
  volume={30},
  number={1},
  pages={85--105},
  year={1998},
  publisher={Annual Reviews 4139 El Camino Way, PO Box 10139, Palo Alto, CA 94303-0139, USA}
}

@article{noh2022atomization,
  title={Atomization characteristics of slurry fuels using a pressure swirl atomizer},
  author={Noh, Kwanyoung and Kim, Hyungmin and Kim, Sanghoon and Song, Soonho},
  journal={Journal of Non-Newtonian Fluid Mechanics},
  volume={304},
  pages={104794},
  year={2022},
  publisher={Elsevier}
}

@article{xu2014study,
  title={Study of droplet formation process during drop-on-demand inkjetting of living cell-laden bioink},
  author={Xu, Changxue and Zhang, Meng and Huang, Yong and Ogale, Amod and Fu, Jianzhong and Markwald, Roger R},
  journal={Langmuir},
  volume={30},
  number={30},
  pages={9130--9138},
  year={2014},
  publisher={ACS Publications}
}

@article{thievenaz2022onset,
  title={The onset of heterogeneity in the pinch-off of suspension drops},
  author={Thi{\'e}venaz, Virgile and Sauret, Alban},
  journal={Proceedings of the National Academy of Sciences},
  volume={119},
  number={13},
  pages={e2120893119},
  year={2022},
  publisher={National Acad Sciences}
}

@article{miskin2012droplet,
  title={Droplet formation and scaling in dense suspensions},
  author={Miskin, Marc Z and Jaeger, Heinrich M},
  journal={Proceedings of the National Academy of Sciences},
  volume={109},
  number={12},
  pages={4389--4394},
  year={2012},
  publisher={National Acad Sciences}
}

@article{basaran2013nonstandard,
  title={Nonstandard inkjets},
  author={Basaran, Osman A and Gao, Haijing and Bhat, Pradeep P},
  journal={Annual Review of Fluid Mechanics},
  volume={45},
  pages={85--113},
  year={2013},
  publisher={Annual Reviews}
}

@article{chateau2018pinch,
  title={Pinch-off of a viscous suspension thread},
  author={Ch{\^a}teau, Joris and Guazzelli, {\'E}lisabeth and Lhuissier, Henri},
  journal={Journal of Fluid Mechanics},
  volume={852},
  pages={178--198},
  year={2018},
  publisher={Cambridge University Press}
}

@article{gvozdyakov2021improvement,
  title={Improvement of atomization characteristics of coal-water slurries},
  author={Gvozdyakov, Dmitry and Zenkov, Andrey},
  journal={Energy},
  volume={230},
  pages={120900},
  year={2021},
  publisher={Elsevier}
}

@Article{furbank2004experimental,
  Title                    = {An experimental study of pArticle effects on drop formation},
  Author                   = {Furbank, Roy J and Morris, Jeffrey F},
  Journal                  = {Physics of Fluids},
  Year                     = {2004},
  Number                   = {5},
  Pages                    = {1777--1790},
  Volume                   = {16},
  Publisher                = {AIP}
}

@Article{furbank2007pendant,
  Title={Pendant drop thread dynamics of particle-laden liquids},
   Author                   = {Furbank, roy J and Morris, Jeffrey F},
  Journal={International journal of multiphase flow},
  Volume={33},
  Number={4},
  Pages={448--468},
  Year={2007},
  Publisher={Elsevier}
}

@article{umemura2014self,
  title={Self-destabilizing loop observed in a jetting-to-dripping transition},
  author={Umemura, Akira and Osaka, Jun},
  journal={Journal of fluid mechanics},
  volume={752},
  pages={184--218},
  year={2014},
  publisher={Cambridge University Press}
}

@Article{ambravaneswaran2000theoretical,
  title={Theoretical analysis of a dripping faucet},
  author={Ambravaneswaran, Bala and Phillips, Scott D and Basaran, Osman A},
  journal={Physical review letters},
  volume={85},
  number={25},
  pages={5332},
  year={2000},
  publisher={APS}
}

@Article{coullet2005hydrodynamical,
  title={Hydrodynamical models for the chaotic dripping faucet},
  author={Coullet, Pierre and Mahadevan, L and Riera, CS},
  journal={Journal of Fluid Mechanics},
  volume={526},
  pages={1--17},
  year={2005},
  publisher={Cambridge University Press}
}

@article{eigenbrod2020effective,
  title={The effective shear and dilatational viscosities of a particle-laden interface in the dilute limit},
  author={Eigenbrod, Michael and Hardt, Steffen},
  journal={Journal of Fluid Mechanics},
  volume={903},
  pages={A26},
  year={2020},
  publisher={Cambridge University Press}
}

@article{safronov2021fast,
  title={Fast waves development initiated by oscillations of a recoiling liquid filament in a viscous fluid jet},
  author={Safronov, AA and Koroteev, AA and Filatov, NI and Bondareva, NV},
  journal={Thermophysics and Aeromechanics},
  volume={28},
  pages={237--245},
  year={2021},
  publisher={Springer}
}

@article{maron1956application,
  title={Application of ree-eyring generalized flow theory to suspensions of spherical particles},
  author={Maron, Samuel H and Pierce, Percy E},
  journal={Journal of colloid science},
  volume={11},
  number={1},
  pages={80--95},
  year={1956},
  publisher={Elsevier}
}

@Article{eggers2008physics,
  title={Physics of liquid jets},
  author={Eggers, Jens and Villermaux, Emmanuel},
  journal={Reports on progress in physics},
  volume={71},
  number={3},
  pages={036601},
  year={2008},
  publisher={IOP Publishing}
}

@Article{ambravaneswaran2004dripping,
  title={Dripping-jetting transitions in a dripping faucet},
  author={Ambravaneswaran, Bala and Subramani, Hariprasad J and Phillips, Scott D and Basaran, Osman A},
  journal={Physical review letters},
  volume={93},
  number={3},
  pages={034501},
  year={2004},
  publisher={APS}
}

@Article{subramani2006simplicity,
  title={Simplicity and complexity in a dripping faucet},
  author={Subramani, Hariprasad J and Yeoh, Hak Koon and Suryo, Ronald and Xu, Qi and Ambravaneswaran, Bala and Basaran, Osman A},
  journal={Physics of fluids},
  volume={18},
  number={3},
  pages={032106},
  year={2006},
  publisher={AIP}
}

@article{sartorelli1994crisis,
  title={Crisis and intermittence in a leaky-faucet experiment},
  author={Sartorelli, JC and Gon{\c{c}}alves, WM and Pinto, RD},
  journal={Physical Review E},
  volume={49},
  number={5},
  pages={3963},
  year={1994},
  publisher={APS}
}

@article{kiyono1999dripping,
  title={Dripping faucet dynamics by an improved mass-spring model},
  author={Kiyono, Ken and Fuchikami, Nobuko},
  journal={Journal of the Physical Society of Japan},
  volume={68},
  number={10},
  pages={3259--3270},
  year={1999},
  publisher={The Physical Society of Japan}
}

@Article{bertrand2012dynamics,
  title={Dynamics of drop formation in granular suspensions: the role of volume fraction},
  author={Bertrand, T and Bonnoit, C and Cl{\'e}ment, E and Lindner, A},
  journal={Granular Matter},
  volume={14},
  number={2},
  pages={169--174},
  year={2012},
  publisher={Springer}
}

@Article{bonnoit2012accelerated,
  title={Accelerated drop detachment in granular suspensions},
  author={Bonnoit, Claire and Bertrand, Thibault and Cl{\'e}ment, Eric and Lindner, Anke},
  journal={Physics of Fluids},
  volume={24},
  number={4},
  pages={043304},
  year={2012},
  publisher={AIP}
}

@Article{lindner2015single,
  title={Single pArticles accelerate final stages of capillary break-up},
  author={Lindner, Anke and Fiscina, Jorge Eduardo and Wagner, Christian},
  journal={EPL (Europhysics Letters)},
  volume={110},
  number={6},
  pages={64002},
  year={2015},
  publisher={IOP Publishing}
}

@article{hoepffner2013recoil,
  title={Recoil of a liquid filament: escape from pinch-off through creation of a vortex ring},
  author={Hoepffner, J{\'e}r{\^o}me and Par{\'e}, Gouns{\'e}ti},
  journal={Journal of fluid mechanics},
  volume={734},
  pages={183--197},
  year={2013},
  publisher={Cambridge University Press}
}

@Article{zhao2015inhomogeneity,
  title={Inhomogeneity in breakup of suspensions},
  author={Zhao, Hui and Liu, Hai-Feng and Xu, Jian-Liang and Li, Wei-Feng and Lin, Kuang-Fei},
  journal={Physics of Fluids},
  volume={27},
  number={6},
  pages={063303},
  year={2015},
  publisher={AIP Publishing}
}

@Article{mathues2015capillary,
  title={Capillary breakup of suspensions near pinch-off},
  author={Mathues, Wouter and McIlroy, Claire and Harlen, Oliver G and Clasen, Christian},
  journal={Physics of Fluids},
  volume={27},
  number={9},
  pages={093301},
  year={2015},
  publisher={AIP Publishing}
}
 \end{document}